\documentclass[manuscript,screen]{acmart}
\AtBeginDocument{
  }

\usepackage{todonotes}
\usepackage{svg}
\usepackage{xcolor}
\usepackage{tcolorbox}

\newcommand{\longonly}[1]{}

\begin{document}

\title{Are Researchers Being Replaced by Artificial Intelligence?}

\author{Angelo A. Salatino}
\email{angelo.salatino@open.ac.uk}
\orcid{0000-0002-4763-3943}
\affiliation{
  \institution{Knowledge Media Institute, The Open University}
  \city{Milton Keynes}
  \country{UK}
}
\author{Ansgar Scherp}
\email{ansgar.scherp@uni-ulm.de}
\orcid{0000-0002-2653-9245}
\affiliation{
  \institution{Ulm University}
  \city{Ulm}
  \country{Germany}
}
\author{Christin Katharina Kreutz}
\email{ckreutz@acm.org}
\orcid{0000-0002-5075-7699}
\affiliation{
  \institution{TH Mittelhessen – University of Applied Sciences}
  \city{Gießen}
  \country{Germany}
}
\author{Sahar Vahdati}
\email{Sahar.Vahdati@tib.eu}
\orcid{0000-0002-7171-169X}
\affiliation{
  \institution{TIB – Leibniz Information Centre for Science and Technology \& Leibniz University of Hannover}
  \city{Hannover}
  \country{Germany}
}

\sloppy

\newcommand\aas[1]{\colorlet{saved}{.}\color{olive}{#1 }\color{saved}}

\begin{abstract}
A Nature survey~\cite{2023Natur.621..672V} from 2023 involving 1,600 researchers shows that scientists are ``concerned, as well as excited, by the increasing use of artificial-intelligence tools in research.''
This tension frames our central question: Are researchers being replaced by artificial intelligence? We argue that replacement is already underway—not as disappearance, but as a shift from researcher-as-creator to researcher-as-curator. As AI agents increasingly generate hypotheses, papers, and reviews, humans risk retaining responsibility while losing intellectual ownership. This article examines how AI is reshaping the scientific lifecycle and exposes the deeper danger: not that AI will fail to do science, but that humans may stop truly understanding it.
\end{abstract}


\maketitle

\tcbset{
  fancybox/.style={
    colback=blue!5,
    colframe=blue!20,
    coltitle=black,
    fonttitle=\bfseries,
    boxrule=1pt,
    arc=6pt,
    left=8pt,
    right=8pt,
    top=8pt,
    bottom=8pt,
    title=#1
  }
}

\begin{tcolorbox}[fancybox={Join the Discussion}]
We invite the community to comment on this working document. 
What do you think, is our profession at risk of being replaced?
What would be proper countermeasures or means for true co-existence of AI and human researchers?

We are happy to hear from you, please sent us an email.
\end{tcolorbox}

\section{Introduction}

Research is advancing human knowledge. 
This knowledge accumulates, allowing us to prosper and thrive because we stand on the shoulders of other researchers. 
As a consequence, research needs to be FAIR, i.\,e., findable, (ideally) freely accessible, interoperable, and reusable\longonly{~\cite{wilkinson2016fair}}.\footnote{FAIR Principles --- \url{https://www.go-fair.org/fair-principles/}.}
So what changes when AI and agents come into play?
We discuss this question by looking back to October 2025. 
A group of researchers at Stanford experimented with using AI agents for science.\footnote{Open Conference of AI Agents for Science 2025 --- \url{https://agents4science.stanford.edu/}.}
This was considered the first conference where AI is not only allowed to be used, but AI was actually required to serve as the primary authors of research papers as well as reviewers.

This raises a serious question:
are researchers being replaced by AI?
The deeper risk emerging from this development is a shift of researchers' roles from \textit{researcher-as-creator} to \textit{researcher-as-curator}, where the intellectual work is outsourced but the responsibility remains with the human.
We are currently witnessing this transition in what we have known as \textit{research}. 
So, we reflect on our guiding question by asking: what forms of AI support for research already exist? How do these tools align with the scientific lifecycle? And how are they reshaping our professional roles?

In general, the scientific lifecycle is a ``structured, iterative framework that sets the ground for the development of empirical knowledge''~\cite{role-of-ai-in-scientific-research-EU-document}.
It consists of activities regarding ideation, planning, execution, creation, and sharing.
This suggests that \textit{replacement} is not an all-or-nothing outcome but a redistribution of agency across this lifecycle.
AI can execute many transferable steps at scale, while researchers must retain scientific judgment, governance, and accountability.

\section{AI-based Research Tools}
\label{sec:ai-in-research}

AI-based tools for research are a new reality, available and used for virtually all research tasks today.
Across research domains, AI-based support can be viewed in terms of 
generic tasks, which recur essentially in every research field, and domain-specific tasks that depend on a community’s particular concepts, data types, tools, and standards.
\longonly{\textit{Generic research tasks} include activities such as literature discovery and review, summarization and synthesis, research-question refinement, drafting and revision, project coordination, and basic data handling. 
These are workflows that are largely transferable across disciplines and therefore well-suited to reusable, general-purpose agents. 
In contrast, \textit{domain-specific tasks} are tightly coupled to disciplinary practice (e.\,g., the methods, ontologies, instruments, paradigms, simulation pipelines, or evaluation criteria used in that field), and therefore benefit from specialized agents that encode domain knowledge and interface directly with field-specific resources. 
}
Examples of generic AI tools, \textbf{applicable across research fields}, are  
\href{https://github.com/yiren-liu/personaflow}{PersonaFlow} for ideation, 
\href{https://github.com/yiren-liu/coquest}{CoQuest} for composing research questions, 
\href{https://github.com/peterjhwang/slr-helper}{SLR Helper} for literature reviews,  
\href{https://github.com/facebookresearch/SelfCite}{SelfCite} for citation generation, 
\href{https://github.com/Northwestern-CSSI/SciSciGPT}{SciSciGPT} for developing or implementing methods, 
\href{https://ai.scidb.cn/en}{ScienceDB AI} for searching for data, 
\href{https://datadreamer.dev/docs/latest/index.html}{DataDreamer} for synthetically generating data, 
\href{https://github.com/VichyTong/dango}{Dango} for wrangling data,
\href{https://github.com/snap-stanford/MLAgentBench/}{MLAgentBench} for evaluating methods, 
\href{https://github.com/zyzisastudyreallyhardguy/LLM4SD}{LLM4SD} for deriving conclusions from evaluations, 
\href{https://paperreview.ai}{Stanford Agentic Reviewer} for reviewing research, 
and on and on.
There is a multiplicity of \textbf{domain-specific tools} such as 
\href{https://ape.socialcatalystlab.org/}{Project APE} for policy evaluation in economic research, 
\href{https://www.qedscience.com}{q.e.d.} for identifying gaps in papers in the life sciences, 
\href{https://github.com/Anonymous-2025-Repr/LLM-DS-Reproducibility}{AIRepr} for evaluating the reproducibility of LLMs in data science or 
\href{https://github.com/CenterForOpenScience/llm-benchmarking}{ReplicatorBench} for benchmarking LLM agents for replicability in social and behavioral sciences.
Additionally, integrated systems are emerging that can handle a \textbf{broader swath of the research lifecycle}: 
\href{https://github.com/jimmc414/Kosmos}{KOSMOS} can be orchestrated across domains for literature synthesis, hypothesis generation, and data interpretation, 
\href{https://allenai.org/asta/agents}{Asta agents by Ai2} supports literature search, summarization, and data management,
\href{https://scispace.com}{SciSpace} supports the complete literature work in the scientific lifecycle, and
\href{https://manus.im/}{Manus} automatically replicates research papers from a single prompt across research domains.
\href{https://github.com/SakanaAI/AI-Scientist-v2}{Sakana.ai} or 
\href{https://aissistant.tib.eu/}{TIB AIssistant}
push these boundaries even further by automating the end-to-end production of entire scientific papers. 
\href{https://www.wingrants.ai}{WinGrants AI} writes full grant proposals at the EU level.
These advancements signal a shift from AI as a task-specific assistant to a more autonomous collaborator capable of executing complex, multi-step workflows. 

The AI-based research ecosystem is rapidly expanding. Major funders are backing automated discovery,\footnote{AI Scientist --- \url{https://www.aria.org.uk/ai-scientist/}.} while the community focuses on design, evaluation, and governance.\footnote{Special Issue on AI-Enabled Scientific Research: Tools and Governance --- \url{https://dl.acm.org/pb-assets/static_journal_pages/tmis/pdf/ACM-TMIS-CFP-AI4Science-1770320480470.pdf}.} This shift raises critical questions about \textit{to what extent researchers should use AI tools} and \textit{whether human researchers are becoming obsolete}~\cite{10.1145/3726302.3730057}.

\section{Risks and Challenges}
\label{sec:challenges}

We observe a paradox: 
integrating AI into the scientific lifecycle accelerates productivity and is a \emph{welcome support in alleviating repetitive, time-consuming tasks}.
At the same time, it also \emph{threatens to decouple human researchers from the intellectual heavy lifting} that defines scientific rigor. 

As AI practitioners, we advocate for the field’s continued evolution while offering a critical examination of the risks these tools pose to traditional scientific standards.
These challenges outlined below represent an intentionally exaggerated position and thus reflect the upper bound of potential risks and challenges of using AI in research. 

\subsection{Ethical Concerns}
\label{sec:ethics}

If an AI generates the hypothesis, conducts the literature review, and drafts the manuscript, the human's role shifts \emph{from creator to curator}.
In the case of the Open Conference of AI Agents for Science 2025, humans have been replaced completely, being relegated to observers.
So, if AI performs the heavy lifting of critical synthesis, what remains of the human researcher's value?

Replacing researchers with AI is not merely about losing a job, but about losing the intellectual friction that defines the scientific identity. 
Who maintains the authority and oversight to serve as an instance of validation?
Is it even necessary to ensure the validity of knowledge or methods of AI-generated research as long as it produces \emph{positive effects}? 
Are we willing to trade human involvement in science for potentially vast scientific breakthroughs by AI that humans no longer understand, such as a cure for cancer that cannot be explained but verified via animal and human testing? This view is already employed in medicine, where mechanisms of action are not fully understood, e.\,g., \cite{modafinil}, yet drugs are used to treat diseases.
As a counter-movement, voices argue to preserve humans in science\footnote{Please, don't automate science! --- \url{https://togelius.blogspot.com/2025/12/please-dont-automate-science.html}.} who ``prefer the joy of doing science to finding a cure for cancer''.

Current institutional frameworks, such as the COPE guidelines\footnote{COPE guidelines ---  \url{https://publicationethics.org/guidance/cope-position/authorship-and-ai-tools}.}, maintain that AI cannot be credited as an author as it cannot hold ethical or legal responsibility. 
However, this creates a conceptual paradox: 
if an AI provides the core intellectual contribution, the human signatory effectively becomes a ghostwriter or a proxy for the machine.

If humans are sidelined, for whom are scientific papers written? 
Should knowledge be represented differently to be even more suitable to be fed to other AI tools? 
Preprint archives today, e.\,g., arXiv\footnote{Formal Requirements --- \url{https://info.arxiv.org/help/policies/format_requirements.html}.}, already provide their content in machine-readable format such as \LaTeX.
Why should we stop here and hold on to an obsolete 
format which no longer serves any obvious purpose to advance science?

How can we measure and safeguard the distinct value of the human researcher in an AI-augmented landscape? Without a clear answer, the transition from creator to curator may eventually lead to a total obsolescence of the expertise required to govern these systems. 
The current generation of researchers will determine the future of the field, deciding not only whom research should serve, but whether science will settle for knowing things \emph{work somehow} but not exactly \emph{how}.

\subsection{Legal Constraints}
\label{sec:legal_constraints}

The legal landscape is a patchwork of reactive policies that researchers must navigate. 
While AI is increasingly accepted for composing papers, it cannot be given authorship,\footnote{See, e.\,g., the \href{https://www.acm.org/publications/policies/frequently-asked-questions}{ACM GenAI policy}, the \href{https://journals.ieeeauthorcenter.ieee.org/become-an-ieee-journal-author/publishing-ethics/guidelines-and-policies/submission-and-peer-review-policies/}{IEEE one}, or the \href{https://publicationethics.org/guidance/cope-position/authorship-and-ai-tools}{COPE guidelines}.} as it cannot take legal responsibility of the submitted work.
However, the lack of accountability creates a vacuum: mishaps, such as the hundreds of AI-hallucinated references found in over 50 accepted NeurIPS 2025 papers, often draw criticism but currently lack tangible consequences for the human authors.
In response, conferences like SIGIR 2026 have begun asking reviewers to flag the undue use of AI. This shift, however, effectively offloads the burden of enforcement onto the peer-review process, creating an asymmetric ecosystem where the sheer volume of AI-augmented submissions threatens to overwhelm the finite capacity of human reviewers. 

When it comes to improving the quality and readability of human-written peer reviews, the use of AI is oftentimes allowed, see, e.\,g., the ACM Peer Review policy.\footnote{ACM Peer Review Policy --- \url{https://www.acm.org/publications/policies/peer-review}.}
However, uploading submissions to AI tools in peer review for conferences and funding agencies remains largely prohibited, with exception of the ``la Caixa'' Foundation.\footnote{\url{https://caixaresearch.org/documents/d/caixaresearch/health-research-call-eligibility-evaluation-guidelines-2026-pdf}}
These bans are intended to protect the confidentiality of unpublished manuscripts and ensure human accountability. 
In addition, using automated reviewing tools does not require a human-based peer-review process: 
authors of a research paper can use such review tools themselves at any point in the writing process.
Dedicated tools for automated paper review are already a reality: the \href{https://paperreview.ai}{Stanford Agentic Reviewer} demonstrated an immense demand for automated reviews, having processed over $20,000$ papers in its inaugural week\footnote{Andrew Ng's post on LinkedIn --- \url{https://www.linkedin.com/posts/andrewyng_releasing-a-new-agentic-reviewer-for-research-activity-7401399616883380224-swcD/}.}. 
This renders their use by actual human reviewers in a regular review process pointless.
Nevertheless, the use of AI has already compromised the peer-review process. Recent analysis suggests that 21\% of reviews at ICLR may be AI-generated~\cite {repec:nat:nature:v:648:y:2025:i:8093:d:10.1038_d41586-025-03506-6}.

\subsection{Skilling Challenges}
\label{sec:skilling_challenges}

What would happen to our skills and those of future generations? 
One important and perhaps the most enduring challenge concerns the erosion of cognitive competencies in the next generation of scholars. 
As AI agents increasingly automate the foundational stages of research, they may remove the educational friction that traditionally builds expertise.
This uncritical reliance risks devaluing human judgment and may leave researchers unable to identify technical errors, spot hallucinations, or conceptualize the paradigm-shifting anomalies that lie beyond a model's training data.

If scholar training shifts from original thought to mere output curation, the scientific community may face a profound competency gap, where a generation can operate the tools of science but lacks the fundamental logic to question them. Ultimately, this atrophy of skill undermines the peer-review process, leaving researchers no choice but to rely on AI-generated reviews instead of human-written ones, further decoupling human intellect from the scientific lifecycle.

\subsection{Scholarly Communication and Validation at Scale}

If \emph{replacement} is understood as a redistribution of agency, the critical question is where accountability rests when AI output scales faster than scrutiny. As AI shifts the bottleneck from producing drafts to evaluating claims, scholarly communication and research integrity will determine whether increased automation yields progress or merely accelerates noise. Beyond legal frameworks, the core challenge will be who will read, filter, reproduce, and integrate an exploding literature? 
The risk is that publication counts will decouple from scientific progress, exacerbated by misaligned incentives where papers serve as essential proxies for academic reputation and funding~\cite{doi:10.1073/pnas.2401231121}. 

As AI reduces the cost of producing manuscripts and feedback, scarce review resources are further strained: careful peer review and synthesis do not scale automatically with generation. While agentic systems might help absorb this validation burden, such as through automated artifact evaluation~\cite{baek2026artisan}, the imbalance is already visible in fast-growing venues\footnote{For example, the official ICML conference account reported 33{,}540 active abstracts immediately after the ICML 2026 abstract deadline (Jan 23, 2026, Anywhere-on-Earth): \url{https://x.com/icmlconf/status/2015105119289803093}.} where submission volumes stretch community attention to the limit.

This pressure is a problem of scientific integrity, not just logistics.
When the cost of producing plausible papers, figures, and analyses approaches zero, incentives shift toward quantity, strategic framing, and metric-driven gaming. 
The burden shifts downstream: the permanent scientific record becomes easier to pollute with low-quality, manipulated, or fabricated results, and subsequent scholarship, and future AI systems trained on the literature, inherit that noise. 
The result is a feedback loop in which unreliable publications degrade the evidence base for both human decisions and automated research pipelines.

\subsection{Technical Challenges}
To assess whether AI replacing researchers \textit{could technically fail}, we must separate two questions that are often conflated:
(i) whether AI can automate large parts of the scientific lifecycle, and
(ii) whether it is intelligent enough to generate and validate genuinely novel scientific contributions.
Current systems are increasingly effective at \textit{scaling the scientific lifecycle}, but far less reliable at \textit{ensuring validity and producing novelty}.

Most current systems primarily expand throughput: drafting, summarizing, coding, and coordinating; by \textit{automating parts of the scientific lifecycle}, to maximize what is already known (see Section~\ref{sec:ai-in-research}).
They also largely contribute to lowering the cost of iteration across the research lifecycle.
But these agentic research tools come with the typical downsides of LLMs~\cite{10.1145/3789199}, such as hallucinations, limited explainability, and data pollution.
Data pollution introduces technical failure modes that are easily missed by humans.
End-to-end pipelines can produce polished outputs while masking invalid evaluation caused by benchmark selection, data leakage, metric misuse, and post-hoc selection bias.

However, the stronger replacement claim depends on the capacity for \textit{novelty and skill-acquisition}~\cite{chollet2019measure}, specifically whether AI can reliably propose hypotheses  beyond mere recombination, and subject them to valid evaluation.
A longer-term technical risk is synthetic self-reinforcement.
Increasing reliance on AI-generated manuscripts, code, and summaries as training material may narrow the diversity of future model outputs.
In the extreme, recursive training on generated data can lead to \textit{model collapse}, where distribution tails disappear and outputs become increasingly repetitive, which would be especially damaging for scientific novelty~\cite{zenil2026limits}.

\section{Conclusion}

Coming back to our central question: will AI replace us?
This concerns the future human role in science: will we be ``degraded'' to solely verifying AI-generated research, thereby losing our critical thinking skills?
Will AI begin to ``outsource'' tasks to human researchers? 
Will AI produce a more structured form of research output, relinquishing scientific papers and human-readability?

While this may sound too far-fetched, it is quickly becoming a reality.
For example, platforms like Rent-a-Human\footnote{\url{https://rentahuman.ai/}} already allow AI agents to hire human labor for tasks they cannot perform, e.\,g., attending a meeting with other humans at a conference.
Similarly, aiXiv\footnote{\url{https://aixiv.science/}} serves as an open pre-print server for research authored by humans and AI.
Students in Germany\footnote{Working with text-generating AI systems --- \url{https://www.schulministerium.nrw/system/files/media/document/file/handlungsleitfaden_ki_msb_nrw_230223.pdf}.} are explicitly allowed to use ChatGPT as co-writers (with formal attribution in footnotes); this may foreshadow a future in professional research in which AI-assisted co-authorship, complete with credited prompts, may become standard practice.

Bottom line, there will be a silver lining, or a path of
mutual benefit for agentic AI and humans doing research.
Surely, there has to be a willingness to adapt, to change one's skill set, work routines, and scientific procedures.
By embracing affordable agentic AI as a ``Swiss army knife'' of research, our technology stack, learning methods, and ethical frameworks will evolve. At some point, AI agents might become co-authors, assume responsibility for their \textit{actions}, or even participate as humanoid robots at future conferences.

\paragraph{Acknowledgment}
We are grateful for the discussions and input from the participants of the Dagstuhl-Seminar 25381 ``Open Scholarly Information Systems: Status Quo, Challenges, Opportunities''.
Ansgar Scherp acknowledges that this research is co-funded by the SocEnRep project (No. 551687338) of the DFG, German Research Foundation.

\bibliographystyle{ACM-Reference-Format}
\bibliography{bib}

@String{Computing = "Computing" }

@String{Academic = "Academic Press" }

@inproceedings{10.1145/3726302.3730057,
author = {Gurevych, Iryna},
title = {Please meet AI, Our Dear New Colleague. In other Words: Can Scientists and Machines Truly Cooperate?},
year = {2025},
isbn = {9798400715921},
address = {New York, NY, USA},
url = {https://doi.org/10.1145/3726302.3730057},
doi = {10.1145/3726302.3730057},
booktitle = {SIGIR '25},
pages = {3–4},
numpages = {2},
}

@book{role-of-ai-in-scientific-research-EU-document,
author = {Purificato, E. and Bili, D. and Jungnickel, R. and Ruiz-Serra, V. and Fabiani, J. and Abendroth-Dias, K. and Fernández-Llorca, D. and Gómez, E.},
title = {The role of artificial intelligence in scientific research – A science for policy, European perspective},
publisher = {Publications Office of the European Union},
year = {2025},
url = {https://data.europa.eu/doi/10.2760/7217497}}

@article{modafinil,
author = {Gerrard, Paul and Malcolm, Robert},
year = {2007},
month = {07},
pages = {349-64},
title = {Mechanism of modafinil: A review of current research},
volume = {3},
journal = {Neuropsychiatric disease and treatment}
}

@Article{repec:nat:nature:v:648:y:2025:i:8093:d:10.1038_d41586-025-03506-6,
journal={Nature},
author={Miryam Naddaf},
title={Major AI conference flooded with peer reviews written fully by AI},
year={2025},
month={December},
pages={256-257},
volume={648},
number={8093},
doi={10.1038/d41586-025-03506-6},
url={https://ideas.repec.org/a/nat/nature/v648y2025i8093d10.1038_d41586-025-03506-6.html},
}

@article{doi:10.1073/pnas.2401231121,
author = {Jennifer S. Trueblood  and David B. Allison  and Sarahanne M. Field  and Ayelet Fishbach  and Stefan D. M. Gaillard  and Gerd Gigerenzer  and William R. Holmes  and Stephan Lewandowsky  and Dora Matzke  and Mary C. Murphy  and Sebastian Musslick  and Vencislav Popov  and Adina L. Roskies  and Judith ter Schure  and Andrei R. Teodorescu },
title = {The misalignment of incentives in academic publishing and implications for journal reform},
journal = {Proceedings of the National Academy of Sciences},
volume = {122},
number = {5},
pages = {e2401231121},
year = {2025},
doi = {10.1073/pnas.2401231121},
URL = {https://www.pnas.org/doi/abs/10.1073/pnas.2401231121}}

@article{chollet2019measure,
  title={On the measure of intelligence},
  author={Chollet, Fran{\c{c}}ois},
  journal={arXiv preprint arXiv:1911.01547},
  doi={10.48550/arXiv.1911.01547},
  year={2019}
}

@article{zenil2026limits,
  title={On the Limits of Self-Improving in LLMs and Why AGI, ASI and the Singularity Are Not Near Without Symbolic Model Synthesis},
  author={Zenil, Hector},
  journal={arXiv preprint arXiv:2601.05280},
  doi={10.48550/arXiv.2601.05280},
  year={2026}
}

@article{wilkinson2016fair,
  author = {Wilkinson, Mark D and Dumontier, Michel and Aalbersberg, IJsbrand Jan and Appleton, Gabrielle and Axton, Myles and Baak, Arie and Blomberg, Niklas and Boiten, Jan-Willem and da Silva Santos, Luiz Bonino and Bourne, Philip E and others},
  journal = {Scientific data},
  publisher = {Nature Publishing Group},
  title = {The FAIR Guiding Principles for scientific data management and stewardship},
  volume = 3,
  year = 2016
}

@article{10.1145/3789199,
author = {Denning, Peter J.},
title = {The Conundrum of LLMs},
year = {2026},
issue_date = {March 2026},
publisher = {Association for Computing Machinery},
address = {New York, NY, USA},
volume = {69},
number = {3},
issn = {0001-0782},
url = {https://doi.org/10.1145/3789199},
doi = {10.1145/3789199},
journal = {Commun. ACM},
month = feb,
pages = {31–34},
numpages = {4}
}

@article{baek2026artisan,
  title={Artisan: Agentic Artifact Evaluation},
  author={Baek, Doehyun and Pradel, Michael},
  journal={arXiv preprint arXiv:2602.10046},
  year={2026}
}

@article{2023Natur.621..672V,
  title={AI and science: what 1,600 researchers think},
  author={Van Noorden, Richard and Perkel, Jeffrey M},
  journal={Nature},
  volume={621},
  number={7980},
  pages={672--675},
  year={2023}
}

\end{document}